\documentclass[12pt]{article}
\RequirePackage[OT1]{fontenc}
\RequirePackage{amsthm,amsmath}
\usepackage{accents}
\usepackage{breqn}
\usepackage{soul}
\usepackage[numbers,sort&compress]{natbib}
\usepackage{soul,xcolor}
\usepackage{hyperref}
\usepackage{cleveref}
\usepackage{amsfonts}
\usepackage{amssymb}
\usepackage{graphicx}
\usepackage{graphics}
\usepackage{setspace}
\usepackage{physics}

\newtheorem{definition}{Definition}[section]

\numberwithin{equation}{section}
\def\b1{{\mathbf 1}}
\def\b0{{\mathbf 0}}

\def\bA{{\boldsymbol A}}
\def\bB{{\boldsymbol B}}

\def\bC{{\boldsymbol C}}

\def\bp{{\boldsymbol p}}

\def\bx{{\boldsymbol x}}

\def\bS{{\boldsymbol S}}

\def\cE{{\mathcal E}}

\def\cI{{\mathcal I}}
\def\cJ{{\mathcal J}}

\def\bbI{{\mathbb I}}

\begin{document}

\title{A predictive solution of the Einstein, \\
Podolski and Rosen's  paradox}
\author{Henryk Gzyl\\
Center for  Finance, IESA School of Management.\\ Caracas, Venezuela.\\
henryk.gzyl@iesa.edu.ve}
\date{}
\maketitle
\setlength{\textwidth}{4in}
\vskip 1 truecm
\baselineskip=1.5 \baselineskip \setlength{\textwidth}{6in}
%%%%%%%%%%%%%%%%%%%%%%%%%%%%%%%%%%%%%%%%%%%
\begin{abstract}
In this work, we examine the paradox proposed by Einstein, Podolsky, and Rosen (EPR). They argued that since one may know the exact momentum of a particle without measurement and subsequently measure its position, a contradiction with the Heisenberg uncertainty principle arises. 

We demonstrate that there is no paradox by two equivalent approaches: first, by computing the quantum conditional expectation to make predictions after  a measurement; and second, using the von Neumann post-measurement state. We establish the equivalence between these two methods. In both cases the predictor is an operator valued function of the observables being measured. This ensures that no violation of the Heisenberg uncertainty principle occurs.
\end{abstract}

\noindent \textbf{Keywords}: EPR paradox, prediction, conditional probabilities as predictors. \\
\small{\tableofcontents}

\section{Introduction and Preliminaries}
In an interesting paper Einstein, Podolsky and Rosen (\cite{EPR}), presented a criticism of the intrinsic randomness of quantum mechanics, and forced the physics community to deepen the understanding of the probabilistic nature of quantum theory.

The core of the EPR argument goes as follows: They assume that the system consists of two, possibly interacting particles, that are in some initial entangled state, which after some time the system evolves into a system of two non-interacting particles, of zero total momentum. As the total momentum is given, they assert, correctly, that if the moment of one particle (the first) is measured, that of the other (the second) is also known without any need of measuring it. According to their argument, since that momentum is known precisely without any measurement, and since the position of the second particle can be measured with arbitrary precision, the Heisenberg uncertainty principle is contradicted.

The effort to overcome the paradox deepened the understanding of quantum mechanics and led to many developments, like Bell inequalities. locality, decoherence, which have already become textbook material. See \cite{J}, \cite{LeB} or \cite{La}. A review was presented in \cite{Ku} and \cite{Ku2}. A compilation of the literature by topic on a yearly basis is posted in \cite{www}. The basic probabilistic tools needed for the argument are detailed in \cite{Pam} and in full detail in \cite{Gu}. For early applications of probabilistic and statistical methods in quantum mechanics, with emphasis on quantum signal analysis, see \cite{He}, \cite{Ho}.

In a two particle system, we may choose as complete system of observables the two momenta $(\hat{\bp}_1,\hat{\bp}_2).$ If we measure the total momentum $\hat{\bp}=\hat{\bp}_1, \hat{\bp}_2,$ then according to the von Neumann measurement theory, the state switches to a state of given total momentum and, if we use the Schr\"{o}dinger representation, any further predictions about the system are to be made with this new state. If we choose to work in the Heisenberg representation, after the measurement, the density matrix changes into a (quantum) conditional density matrix.

We show how these changes in the state vector or in the density matrix, solve the paradox that they formulated.

The remainder of the paper is organized as follows. In Section 2, after briefly recalling EPR' s setup and the statement of the paradox. Then we show how to compute the quantum conditional expectation of the momentum of one of the particles (the first) given the total momentum of the system.  
 The important theoretical issue is that quantum conditional expectations are operator valued. In our case, they are functions of the total momentum, which is distributed according to a probability density that depends on the initial state of the system. This should be taken into account when making further predictions about the system.

In Section 3 we show that predicting with the (von Neumann) post-measurement state coincides with predicting with quantum conditional expectations. This unifies both points of view and further emphasizes the dependence of the predictors on the total momentum of the system. 

In Section 4 we address the paradox proper. It is here that the operator nature of the quantum conditional expected values is essential to solve the paradox posed by EPR. 

In Section 5 we consider an explicit example in which the state is such that the density matrix is an entangled Gaussian state in the momentum representation. Then we compute the conditional density matrix as in Section 2, and use it to verify that the prediction of the momentum of the first particle depends on the total momentum.

In Section 6 we sum up and leave some computations of potential interest for the Appendix.

%%%%%%%%%%%%%%%%%%%%%%%%%%%%%%%%%%%%%%%%%%%%%%%%%%%%%%%%%%%
%%%%%%%%%%%%%%%%%%%%%%%%%%%%%%%%%%%%%%%%%%%%%%%%%%%%%%%%%
\section{Prediction with prior observation}
To motivate our choice of setup, we first recall the original proposal by Einstein, Podolsky, and Rosen. In their seminal paper \cite{EPR}, they suggest the following state in the coordinate representation:
\begin{equation}\label{epr1}
\Psi(x_1, x_2) = \int_{-\infty}^{\infty} e^{ip(x_1 - x_2 + x_0)/\hbar} \, dp.
\end{equation}
The intent of this state is to represent a system with zero total momentum. However, a well-known formal difficulty arises: this state is not an element of the Hilbert space of square-integrable functions $L^2(\mathbb{R}^2)$. Consequently, one cannot strictly invoke Born's probabilistic interpretation to compute well-defined probabilities. Using \eqref{epr1} to compute expected values and variances is, from a rigorous standpoint, problematic from the outset. This technicality impinges on the EPR argument, which relies on the precise calculation of variances to demonstrate a purported contradiction of the Heisenberg uncertainty principle.

To establish standard notation, we assume that physical states are vectors in a Hilbert space $\mathcal{H}$. If $\Psi \in \mathcal{H}$ (or $|\Psi\rangle \in \mathcal{H}$) denotes a state in the Schrödinger representation, we use $\rho = |\Psi\rangle\langle\Psi|$ to denote the corresponding density operator. For a Hermitian operator $\mathbf{A}$ describing a physical observable, the expected value is denoted by:
\begin{equation}\label{not1}
\mathcal{E}_{\rho}[\mathbf{A}] = E_{\Psi}[\mathbf{A}] = \text{tr}(\mathbf{A} |\Psi\rangle\langle\Psi|) = \text{tr}(\mathbf{A}\rho) = \langle\Psi|\mathbf{A}|\Psi\rangle.
\end{equation}
This value serves either as a predictor for a future observation of $\mathbf{A}$ or as the statistical report of an experiment measuring $\mathbf{A}$.

In this work, we use the momenta $\hat{\bp}_1$ and $\hat{\bp}_2$ as the complete system of observables. In the momentum representation, the system is described by a wave function $\Psi(p_1, p_2)$ such that $\rho(p_1, p_2) = |\Psi(p_1, p_2)|^2$ is a valid probability density integrating to $1$.

To determine the probability density of observing a specific value $p$ of the total momentum $\hat{\bp} = \hat{\bp}_1 + \hat{\bp}_2$, we consider an arbitrary test function $h(\hat{\bp})$ and compute its expectation:
\begin{equation}
\begin{aligned}
\mathcal{E}_\Psi[h(\hat{\bp})] &= \int h(p_1 + p_2) \rho(p_1, p_2) \, dp_1 dp_2 \\
&= \int h(p) \left( \int \rho(p_1, p - p_1) \, dp_1 \right) dp.
\end{aligned}
\end{equation}
From this, we conclude that the probability density of the total momentum $\hat{\bp}$ in state $\Psi$ is:
\begin{equation}\label{dens1}
\rho_{\hat{\bp}}(p) = \int \rho(p_1, p - p_1) \, dp_1 = \int \rho(p - p_2, p_2) \, dp_2.
\end{equation}
This is the first step in the computation of the density of  $\hat{\bp}_1$ conditioned on $\hat{\bp}$. If the initial state were separable, $\Psi(p_1, p_2) = \Psi_1(p_1)\Psi_2(p_2)$, the result in \eqref{dens1} would reduce to the standard convolution of densities: $\rho_{\hat{\bp}}(p) = \int \rho_1(p_1)\rho_2(p - p_1) \, dp_1$.

%%%%%%%%%%%%%%%%%%%%%%%%%%%%%%%%%%%%%%%%%%%%%%%%%%%%%%%%%%%%%%%%%%%%%%%%%%%%%%%%
\subsection{Quantum Conditional Expectations}
We now recall basic properties of quantum conditional expectations, following the framework in \cite{Gu}. In our specific case, the observables $\hat{\bp}_1, \hat{\bp}_2$, and $\hat{\bp}$ all commute. Thus, the operator-theoretic technicalities simplify significantly, mirroring the classical probabilistic case.

\begin{definition}\label{CE0}
Suppose a system is in state $\Psi \in \mathcal{H}$. Let $F(\hat{\bp}_1)$ be an observable such that $\mathcal{E}_{\Psi}[|F(\hat{\bp}_1)|] < \infty$. The quantum conditional expectation of $F(\hat{\bp}_1)$ given an observation of $\hat{\bp}$ is an operator-valued function of $\hat{\bp}$, denoted $\mathcal{E}_{\Psi}[F(\hat{\bp}_1) | \hat{\bp}]$, which satisfies:
\begin{equation}\label{CE1}
\mathcal{E}_{\Psi}\big[ \mathcal{E}_\Psi[F(\hat{\bp}_1) | \hat{\bp}] G(\hat{\bp}) \big] = \mathcal{E}_\Psi[F(\hat{\bp}_1) G(\hat{\bp})],
\end{equation}
for any bounded $G(\hat{\bp})$. Furthermore, this operator satisfies:
\begin{equation}\label{CE2}
\mathcal{E}_\Psi[K(\hat{\bp}_1, \hat{\bp}) | \hat{\bp}_1, \hat{\bp}] = K(\hat{\bp}_1, \hat{\bp})
\end{equation}
\noindent for any $K(\hat{\bp}_1, \hat{\bp}) .$
\end{definition}

In the classical notation, we write $\mathcal{E}_{\Psi}[F(\hat{\bp}_1) | \hat{\bp} = p]$ to represent the value of the expectation when a measurement of $\hat{\bp}$ yields $p$. The condition \eqref{CE2} means, that observations of $\hat{\bp}$ and $\hat{\bp}_1$ yield $p$ and $p_1$, then for any observable $K(\hat{\bp}_1, \hat{\bp})$ we must have:
\begin{equation}\label{interp1}
\mathcal{E}_\Psi[K(\hat{\bp}_1, \hat{\bp}) | \hat{\bp}_1 = p_1, \hat{\bp} = p] = K(p_1, p).
\end{equation}

Applying \eqref{CE1}, we can derive explicitly the conditional density $\rho_{\hat{\bp}_1 |  \hat{\bp}}(p_1 | p)$ as follows. Consider observables $F$ and $G$ and compute:
\begin{equation}\label{conex1}
\begin{aligned}
\mathcal{E}_\rho[F(\hat{\bp}_1) G(\hat{\bp})] &= \iint F(p_1) G(p_1 + p_2) \rho(p_1, p_2) \, dp_1 dp_2 \\
&= \int G(p) \left( \int F(p_1) \frac{\rho(p_1, p - p_1)}{\rho_{\hat{\bp}}(p)} \, dp_1 \right) \rho_{\hat{\bp}}(p) \, dp.
\end{aligned}
\end{equation}
Therefore, the conditional density $\rho_{\hat{\bp}_1 | \hat{\bp}}$ and the conditional expectation are:
\begin{gather}
\rho_{\hat{\bp}_1 | \hat{\bp}}(p_1 | p) = \frac{\rho(p_1, p - p_1)}{\rho_{\hat{\bp}}(p)}, \label{conden1} \\
\mathcal{E}_\rho[F(\hat{\bp}_1) | \hat{\bp} = p] = \int F(p_1) \frac{\rho(p_1, p - p_1)}{\rho_{\hat{\bp}}(p)} \, dp_1. \label{conex2}
\end{gather}
To  predict $\hat{\bp}_2$ given both $\hat{\bp}$ and $\hat{\bp}_1$,  we take  $K(\hat{\bp}_1, \hat{\bp}) =\hat{\bp}-\hat{\bp}_1,$ Then \eqref{CE2} asserts that:

\begin{equation}\label{conden2}
\mathcal{E}_\Psi[\hat{\bp} - \hat{\bp}_1 | \hat{\bp}_1, \hat{\bp}] =\hat{\bp} - \hat{\bp}_1
\end{equation}

That is, having observed $\hat{\bp}_1$ and $ \hat{\bp},$ the best predictor of $\hat{\bp}_2$ is $\hat{\bp} - \hat{\bp}_1$ with probability $1.$ Apart of confirming to physical intuition, the important thing to stress is that the best predictor after an observation is an operator valued quantity, or that it itself is an observable.

Besides that, observe that there is nothing else to say about the system. The original probability density has been observed at its two independent variables, that is $\Psi(p_1,p-p_1)|^2$ is known at the observed values, and there are no other  variables specifying the system.

%%%%%%%%%%%%%%%%%%%%%%%%%%%%%%%%%%%%%%%%%%%%%%%%%
%%%%%%%%%%%%%%%%%%%%%%%%%%%%%%%%%%%%%%%%%%%%%%%%%%%%
\section{Measurement and conditional probabilities}
Here we establish a correspondence between the von Neumann formalism to describe measurements and the conditional probability approach presented above. We direct the reader to \cite{J} for details about the description of quantum mechanical measurements. We want to describe what happens when one measures the total momentum. The analytical details are due to the fact that the variables are continuous. In the discrete case the steps are similar and straightforward.
Suppose the initial state of the system is $\Psi\rangle$ and let $ \rho=|\Psi\rangle\langle\Psi|$ be the corresponding density matrix. Suppose we measure whether the total momentum $\hat{\bp}$ lies in an interval $\cI(\epsilon)=(p-\epsilon,p+\epsilon)$ about $p.$ Consider the projection operator
\begin{equation}\label{proj1}
\begin{aligned}
P(p,\epsilon)  =  \int \int \chi_{\cI(\epsilon)}(p_1+p_2)|p_1,p_2\rangle\langle p_1,p_2|dp_1dp_2\\
=\int_{p-\epsilon}^{p+\epsilon}\big(\int |p_1,p'-p_1\rangle\langle p_1,p'-p_1|dp_1\big)dp'= \chi_{\cI(\epsilon)}(\hat{\bp}_1+\hat{\bp}_2).
\end{aligned}
\end{equation}

Since $\chi_{\cI(\epsilon)}$ is a binary function assuming values $0$ and $1,$  $P(p,\epsilon)P(p,\epsilon)=P(p,\epsilon).$  Use $\rho=|\Psi\rangle\langle\Psi|$ as usual. Then:
\begin{equation}\label{NF}
\begin{aligned}
\tr\big(P(p,\epsilon)\rho P(p,\epsilon)\big) =\int \int \chi_{\cI(\epsilon)}(p_1+p_2)|\Psi(p_1,p_2)|^2dp_1dp_2\\
=\int_{p-\epsilon}^{p+\epsilon}\big(\int |\Psi(p_1,p'-p_1)|^2dp_1\big)dp'.
\end{aligned}
\end{equation}

In the Schr\"{o}dinger representation, the post measurement state is:
\begin{equation}\label{PMS1}
\Psi_{\epsilon} = \frac{ P(p,\epsilon)\Psi}{\sqrt{\tr\big(P(p,\epsilon)\rho P(p,\epsilon)\big)}}.
\end{equation}
Whereas in the Heisenberg representation we have:
\begin{equation}\label{normal}
\rho^{\epsilon}= \frac{ P(p,\epsilon)\rho P(p,\epsilon)}{\tr\big(P(p,\epsilon)\rho P(p,\epsilon)\big)}.
\end{equation}

If we want to predict the expected value of an observable $F(\hat{\bp}_1)$ in this state, we compute
\begin{equation}\label{ce1}
\cE_{\rho^{\epsilon}}[F(\hat{\bp}_1)] = tr\big(F(\hat{\bp}_1)\rho^{\epsilon}\big).
\end{equation}
The computation of this trace is simple due to the commutativity of the operators involved. It results in:
\begin{equation}\label{ce2}
tr\big(F(\hat{\bp}_1)\rho^{\epsilon}\big).=\frac{\int_{p-\epsilon}^{p+\epsilon}\big(\int F(p_1) |\Psi(p_1,p'-p_1)|^2dp_1\big)dp'.}{\int_{p-\epsilon}^{p+\epsilon}\big(\int |\Psi(p_1,p'-p_1)|^2dp_1\big)dp'.}.
\end{equation}
When the interval about $p$ becomes infinitesimally small, the expected value of $F(\hat{\bp}_1),$ in the post measurement state  obtained when $\hat{\bp}$ is observed to have value $p,$ is:
\begin{equation}\label{measp1}
\lim_{\epsilon\to 0}\cE_{\rho^{\epsilon}}[F(\hat{\bp}_1)]=\lim_{\epsilon\to 0}tr\big(F(\hat{\bp}_1)\rho^{\epsilon}\big)dp_1=
\frac{ \int F(p_1)|\Psi(p_1,p-p_1)|^2}{\int |\Psi(p_1,p-p_1)|^2dp_1}
\end{equation}
In other words, the expected value of the momentum of the first particle with respect to the state obtained from the original state after measuring the total momentum, computed as in \eqref{measp}, coincides with the conditional density \eqref{conden1}. This establishes the equivalence of the two procedures.

Moreover, \eqref{measp1} points to an interesting issue related to the collapse of the wave function after a measurement. 
The limiting procedure carried out in \eqref{measp1}  cannot be  applied to the Schr\"{o}dinger  representation \eqref{PMS1}.  In the Heisenberg representation the collapsed density corresponds to the conditional density \eqref{conden1}. Also, as apparent in \eqref{measp1}, the reduced density is an operator valued density. 

If $G(\hat{\bp})$ is a continuous function of the total momentum $\hat{\bp},$ a similar computation to that in \eqref{ce2} yields
\begin{equation}\label{ce3}
\cE_{\rho^{\epsilon}}[G(\hat{\bp})] = \frac{\int_{p-\epsilon}^{p+\epsilon} G(p')\big(\int |\Psi(p_1,p'-p_1)|^2dp_1\big)dp'.}{\int_{p-\epsilon}^{p+\epsilon}\big(\int |\Psi(p_1,p'-p_1)|^2dp_1\big)dp'.}.
\end{equation}
And clearly
\begin{equation}\label{measp}
\lim_{\epsilon\to 0}\cE_{\rho^{\epsilon}}[G(\hat{\bp})]= G(p).
\end{equation} Therefore, in this state $\rho_{\hat{p}},$ the predicted error for measurement (the variance) is $0.$

Let us suppose that after measuring $\hat{\bp}$ and reducing the state to $\rho^{\epsilon}$ we measure $\hat{\bp}_1.$ Let $\cJ_{\delta}=(p_1-\delta,p_1+\delta),$ and let $P(p_1,\delta)=\chi_{\cJ_{\delta}}(\hat{\bp}_1).$ 

Then, in the Schr\"{o}dinger representation, the post-measurement state is
\begin{equation}\label{PMS2}
\Psi_{\epsilon,\delta}=\frac{P(p_1,\delta)\Psi_{\epsilon}}{\sqrt{tr\big(P(p_1,\delta)\rho^{\epsilon}P(p_1,\delta)\big)}}.
\end{equation}
In the Heisenberg representation, the post-measurement state is:
\begin{equation}\label{red2}
\rho_{\delta,\epsilon} = \frac{P(p_1,\delta)\rho^{\epsilon}P(p_1,\delta)}{tr\big(P(p_1,\delta)\rho^{\epsilon}P(p_1,\delta)\big)}.
\end{equation}
The computation of the traces is direct. Let as in the previous section, $K(\hat{\bp}_1.\hat{\bp}_2)$ denote an observable of interest. Its predicted value in the state $\rho_{\delta}$ is

\begin{equation}\label{measp2}
\tr\big(K(\hat{\bp}_1,\hat{\bp}_2)\rho_{\delta,\epsilon}\big)=\frac{\int_{p_1-\delta}^{p_1+\delta}\int_{p-\epsilon}^{p+\epsilon}K(p^'_1,p^'-p^'_1)\rho_{\psi}(p^'_1,p^'-p^'_1)dp^'dp^'_1}{ \int_{p_1-\delta}^{p_1+\delta}\int_{p-\epsilon}^{p+\epsilon}\rho_{\psi}(p^'_1,p^'-p^'_1)dp^'dp^'_1}.
\end{equation}

If we  predict $\hat{\bp}_2$ in the state \eqref{PMS2}, we take   $K(\hat{\bp}_1,\hat{\bp}_2)=\hat{\bp}_2=\hat{\bp}-\hat{\bp}_1,$  to obtain:

\begin{equation}\label{pred2.1}
\langle\Psi_{\epsilon,\delta}|\hat{\bp}_2|\Psi_{\epsilon,\delta}\rangle = \frac{\int_{p_1-\delta}^{p_1+\delta}\int_{p-\epsilon}^{p+\epsilon}\big(p^'-p^'_1\big)\rho_{\psi}(p^'_1,p^'-p^'_1)dp^'dp^'_1}{ \int_{p_1-\delta}^{p_1+\delta}\int_{p-\epsilon}^{p+\epsilon}\rho_{\psi}(p^'_1,p^'-p^'_1)dp^'dp^'_1}.
\end{equation}
This can be written as a conditional expectation:
\begin{equation}\label{pred2.2}
\langle\Psi_{\epsilon,\delta}|\hat{\bp}_2|\Psi_{\epsilon,\delta}\rangle=\cE_\Psi[\hat{\bp}_2|\hat{\bp}\in\cI(\epsilon),\hat{\bp}_1\in\cJ(\delta)].
\end{equation}

If the measurements of $\hat{\bp}$ and $\hat{\bp}_1$ are made with very high precision we can make $\delta\to 0$ and $\epsilon\to0$ in \eqref{measp2} to obtain:
\begin{equation}\label{measp2}
\lim_{\delta\to0}\lim_{\epsilon\to0}\tr\big(K(\hat{\bp}_1,\hat{\bp}_2)\rho_{\delta}\big)=\cE_\rho[K(\hat{\bp}_1,\hat{\bp}_2)|\hat{\bp}_1=p_1,\hat{\bp}=p] = K(p_1,p-p_1).
\end{equation}

In particular, when $K(\hat{\bp}_1,\hat{\bp}_2)=\hat{\bp}_2=\hat{\bp}-\hat{\bp}_1,$ then 
\begin{equation}\label{measp3}
\lim_{\delta\to0}\lim_{\epsilon\to0}\tr\big(\hat{\bp}_2\rho_{\delta}\big)=\cE_\rho[ \hat{\bp}_2|\hat{\bp}_1=p_1,\hat{\bp}=p] = p-p_1.
\end{equation}
That is, we reobtain \eqref{conden2}.  another way to write \eqref{measp2} is that:

\begin{equation}\label{col}
|\Psi_{\epsilon,\delta}\rangle\langle\Psi_{\epsilon,\delta}|=\rho_{\delta,\epsilon}\;\;\rightarrow\;\;\delta\big(p'_1-p_1\big)\big(p'_2-(p-p_1)\big).
\end{equation}

That is, the density matrix collapses to a point mass at the observation point when the observation is infinitely precise. Note that the limiting procedure in \eqref{measp2}  cannot be carried out in the Schr\"{o}dinger representation \eqref{PMS2} despite the fact that the limiting procedure is well defined in the Heisenberg representation.  The problem is, this density function in \eqref{col} is not the density matrix of any sate in the Hilbert space of states of the system.

To repeat the comment at the end of Section 2, in a system with two degrees of freedom, once the two observables characterizing the state are measured, there is nothing else to observe about the system. 

%%%%%%%%%%%%%%%%%%%%%%%%%%%%%%%%%%%%%%%%%%%%%%%%%%%%
%%%%%%%%%%%%%%%%%%%%%%%%%%%%%%%%%%%%%%%%%%%%%%%%%%%
\section{Solving the EPR paradox}
Recall that, if $\bA,\bS,\bC$ are three Hermitian operators on a Hilbert space such that $[\bA,\bB]=i\hbar\bC,$ and if $\Psi$ is a normalized vector representing a physical state, the the Heisenberg  uncertainty principle asserts that $\Delta_\Psi(\bA)\Delta_\Psi(\bB)\geq|\hbar\langle\Psi,\bC\Psi\rangle|/2.
.$
This is recalled to stress that in the case of the position and momentum uncertainty relationship, the right hand is $|\langle\Psi,\bC\Psi\rangle|$ which is $1$ if $\Psi$ is a normalized state. Otherwise, we may run into contradictions, especially if $\Psi$ has an infinite norm.

In our situation we have three states that are used to make predictions. First, the state $\Psi$ in which the system is prepared, second the post-measurement state after the total momentum is measured, and third, the state after the total momentum momentum and then the momentum of the first particle are measured. 

After the first measurement is made, the statistical nature of the system is characterized by the (quantum) conditional probability density, and there is no problem.  A few more remarks about this case are deferred to the appendix.

Notice that if, after two measurements are made, and the results are not precise, the post-measurement state \eqref{PMS2} is a well defined state. In this state, the possible values of  $\hat{\bp}_1$ and $\hat{\bp}$ lie in the rectangle $\cJ(\delta)\times\cI(\epsilon),$ With respect to the  state  $\Psi_{\epsilon,\delta}$ the position and the moment of the second particle can be predicted, and the errors in the prediction satisfy the Heisenberg uncertainty principle. Explicitly:

\begin{equation}\label{HUP}
\Delta_{\Psi(\epsilon,\delta)}(\hat{\bx}_2)\Delta_{\Psi(\epsilon,\delta)}(\hat{\bp}_2) \geq\frac{ \hbar}{2}|\langle\Psi(\epsilon,\delta),\Psi(\epsilon,\delta)\rangle|=\frac{\hbar}{2}.
\end{equation}

The case considered by EPR corresponds to the case in which two measurements that yield a precise number are made. This limit case corresponds to \eqref{measp3} or, equivalently,  to \eqref{conden2}. 
In the limit as $\epsilon\to0$ and $\delta\to 0,$ an application of \eqref{measp3} shows that $\Delta_{\Psi(\epsilon,\delta)}(\hat{\bp}_2)\to 0$ while the right hand side stays constant. This implies that necessarily $\Delta_{\Psi(\epsilon,\delta)}(\hat{\bx}_2)\to \infty.$  

As pointed out at the end of Section 3, even though taking the limit as $\epsilon\to0$ and $\delta\to 0,$  in \eqref{HUP} is a well defined procedure, the result does not correspond to the evaluation of the variances in a true state of the two particle system. 

Also, since conditional expectations are operator valued quantities, as stated in \eqref{conden2} or \eqref{measp3}, the predictor of $\hat{\bp}_2$ given $\hat{\bp}$ and $\hat{\bp}_1$  is $\hat{\bp}-\hat{\bp}_1,$ is an observable which becomes known when $\hat{\bp}$ and $\hat{\bp}_1$ are measured. The statistical nature of this predictor is still contained in the density $|\Psi(p_1,p-p_1)|^2.$  Since the position operator of the second particle does not commute with $\hat{\bp}-\hat{\bp}_1,$ the Heisenberg uncertainty principle is in force with respect to the initial state for every non zero value of $\epsilon$ and $\delta.$

%%%%%%%%%%%%%%%%%%%%%%%%%%%%%%%%%%%%%%%%%%%%%%%%%
%%%%%%%%%%%%%%%%%%%%%%%%%%%%%

\section{Explicit example}
We consider a two particle system, whose momenta are the complete set of observables characterizing its state, which in the Heisenberg representation is
\begin{equation}\label{state1}
\rho(p_1,p_2) = \frac{1}{2\pi\sigma^2(1-c^2)}\exp\bigg[-\frac{1}{(1-c^2)}\bigg(\frac{(p_1-\mu_1)^2}{\sigma^2}-2c\frac{p_1-\mu_1}{\sigma}\frac{p_2-\mu_2}{\sigma}+\frac{(p_2-\mu_2)^2}{\sigma^2}\bigg)\bigg].
\end{equation}
Above, $\mu_1$ and $\mu_2$ denote the mean values of the momenta (we ran out of letters). The variances of each momentum variable is $\sigma^2$ and that their correlation is $c.$  To simplify the notations, introduce the centered coordinates $\xi_1=p_1-\mu_1$and $\xi_2=p_2-\mu_2.$ If we are going to observe the total momentum, it is convenient to use as coordinates the variables $(\xi,\zeta)$ where $\xi=\xi_1$ (drop the subscript) and $\zeta=\xi+\xi_2.$ The Jacobian of the change of variables is $1,$ and to obtain the density $\rho(\xi,\zeta)$ in the new coordinates it suffices to rearrange the exponent in \eqref{state1}. After some arithmetical manipulations, one can verify that:
$$\frac{1}{\sigma^2(1-c^2)}\bigg( \xi^2 -2c\xi(\zeta-\xi) +(\zeta-\xi)^2\bigg)=\frac{1}{\sigma^2(1-c^2)}\bigg( 2(1+c)\xi^2 -2(1+c)\xi\zeta-\xi) +\zeta^2\bigg).$$
In vector notation this can be written as $(\xi,\zeta)\Sigma^{-1}(\xi,\zeta)^t$ where the matrix $\Sigma^{-1}$ and its inverse are:
\begin{equation}\label{matr}
\Sigma^{-1}=\frac{1}{\sigma^2(1-c^2)}\begin{pmatrix} 2(1+c) & -(1+c)\\ -(1+c)& 1\end{pmatrix}\;;\;\; \Sigma = \begin{pmatrix} \sigma^2 & \sigma^2(1+c)\\ \sigma^2(1+c) & 2\sigma^2(1+c)\end{pmatrix}.
\end{equation}
The matrix $\Sigma$ displays the covariance of $(\xi,\zeta)$ with itself. That is, the covariance between the $\xi=p_1-\mu_1$ and $\zeta=p-\mu$ ($\mu=mu_1+\mu_2$). A standard result in linear prediction theory in the Gaussian case can be invoked to compute \eqref{conex1} in the case when $g(\hat{\bp}_1)=\hat{\bp}_1.$ See \cite{Br} or \cite{LM} for example. In this case the conditional mean and the conditional variance are:
\begin{equation}\label{linpred1}
\begin{aligned}
\cE_{\rho}[\hat{\bp}_1|\hat{\bp}] = \mu_1 + \frac{1}{2}(\hat{\bp}-\mu) = \langle\hat{\bp}_1\rangle_{\hat{\bp}}.\\
\cE[\big(\hat{\bp}_1-\langle\hat{\bp}_1\rangle_{\hat{\bp}}\big)^2|\hat{\bp}]= \frac{\sigma^2}{2}\big(1-c).
\end{aligned}
\end{equation}

In \eqref{linpred1} we exhibit the predictor of the momentum of the first particle, as a function of the total momentum $\hat{\bp}.$  The prediction of $\hat{\bp}-\hat{\bp}_1$ (the momentum of the second particle) is
$$\cE_{\rho}[\hat{\bp}- \hat{\bp}_1|\hat{\bp}] = \hat{\bp}-E_{\rho}[\hat{\bp}_1|\hat{\bp}]=\frac{1}{2}\big(\hat{\bp}-(\mu_2-\mu_1)\big).$$
Again, the conditional prediction of $\hat{\bp}-\hat{\bp}_1,$ given $\hat{\bp}$ and $\hat{\bp}_1,$ is, according to \eqref{CE2},
$$ \cE_{\rho}[\hat{\bp}- \hat{\bp}_1|,\hat{\bp},\hat{\bp}_1]=\hat{\bp}-\hat{\bp}_1.$$

%%%%%%%%%%%%%%%%%%%%%%%%%%%%%%%%%%%%%%%%%%%%%
%%%%%%%%%%%%%%%%%%%%%%%%%%%%%%%%%%%%%%%%%%%%%%%%%%%%%%%%%%%
\section{Final remarks}
We presented two ways of thinking about predicting the momentum of a particle given that the total momentum and the momentum of the other are observed. The theoretical issue to stress is that the quantum conditional expectation is an operator valued quantity. That means that it itself is an observable, which is a function of the measured observables. As such, it has a statistical nature described by the probability implicit in the original state in which the system is prepared. This is essential for verifying the uncertainty principle referring to measurements involving the predictors and other observables of interest. 

We pointed out that a problem appears when specifying point values for the observed quantities. In this case, after specifying the values of a complete system of observables characterizing the state of the system, the statistical nature of the system disappears. 

This issue is standard in classical probability. If the state of a system is characterized by a random variable $X,$ and $F(X)$ is a function of the random outcome $X,$ then $E[F(X)|X]=F(X)$ with probability $1.$ That the probability is $1$ means that, except for a set of zero probability, $E[F(X)|X=x]=F(x).$ This means that if observe $X=x,$ the prediction of $F(X)$ is $F(x).$ This is part of the logic on the prediction process, not the result of any physical action of the observer.

To close, we add, that in classical probability, there exist random variables with infinite mean or with infinite 
variance.  The fact that a phenomenon shows such unpredictability does not make it less real. The time to cross a barrier by a Brownian motion being a typical example of unpredictable (random variable with infinite mean and infinite variance). See \cite{PG}for a proposal of how to deal with such cases.

\section{Appendix: the predictions after the first measurement}
We saw above that if we describe the initial state of the two particle system by a normalized vector, then, if we measure the total momentum, we obtain a proper, normalized state.  In this case, having observed $\hat{\bp},$ the momentum $\hat{\bp}_1$ of the first particle is distributed according to
$\rho_{\hat{\bp}_1| \hat{\bp}}(p_1|\hat{\bp}),$  given in \eqref{conden1}. The predicted value of $\hat{\bp}_1$ with respect to this density is:
\begin{equation}\label{pred1}
\cE_\rho[\hat{\bp}_1|\hat{\bp}] = \int p_1\frac{\rho(p_1,\hat{\bp}-p_1)}{\rho_{\hat{\bp}}(\hat{\bp})} dp_1.
\end{equation}
Since $\hat{\bp}=\hat{\bp}_1+\hat{\bp}_2,$ the predicted value of $\hat{\bp_2}$ is:
\begin{equation}\label{pred2}
\cE_\rho[\hat{\bp}-\hat{\bp}_1|\hat{\bp}] = \hat{\bp} - \int p_1\frac{\rho(p_1,\hat{\bp}-p_1)}{\rho_{\hat{\bp}}(\hat{\bp})} dp_1.
\end{equation}

Clearly, the right hand sides of \eqref{pred1}-\eqref{pred2} are operator valued functions of the total momentum $\hat{\bp},$  whose possible values are distributed according to $\rho_{\hat{\bp}}(p)$ given by \eqref{dens1}.  These predictors have the same error respect to the conditional density \eqref{conden1} given by

\begin{equation}\label{prederr}
\Delta_{\hat{\bp}}(\hat{\bp}_1) =\bigg(\int(p_1-\langle\hat{\bp}_1\rangle_{\hat{\bp}}\big)^2\rho_{\hat{\bp}_1| \hat{\bp}}(p_1|\hat{\bp})dp_1\bigg)^{1/2} =\Delta_{\hat{\bp}}(\hat{\bp}_2) .
\end{equation}

 The reason for the equality of variances is that both predictors differ by a constant (the given value of $\hat{\bp}$). We use $\langle\hat{\bp}_1\rangle_{\hat{\bp}}$ to denote the conditional means in \eqref{pred1}-\eqref{pred2}.

\textbf{Declaration of competing interests:} We have no competing interests to declare.
\textbf{Data availability} There is no data availability issue associated with this submission.
\end{document}